\documentclass[iop,apj,tighten]{emulateapj}
\usepackage{amsmath,amstext}
\usepackage[breaklinks,colorlinks,citecolor=blue,linkcolor=magenta]{hyperref}
\usepackage[all]{hypcap} 

\shorttitle{Freely Expanding X-ray Knots in Kepler's Supernova Remnant}
\shortauthors{Sato \& Hughes}

\begin{document}

\title{Freely Expanding Knots of X-ray Emitting Ejecta in Kepler's
  Supernova Remnant}
\author{Toshiki \textsc{Sato}\altaffilmark{1,2}
    and John P. \textsc{Hughes}\altaffilmark{3,4}
   }

\newcommand{\myemail}{toshiki@astro.isas.jaxa.jp}
\newcommand{\chandra}{{\it Chandra}}

\altaffiltext{1}{Department of Physics, Tokyo Metropolitan University, 1-1 Minami-Osawa, Hachioji, Tokyo 192-0397}
\altaffiltext{2}{Department of High Energy Astrophysics, Institute of Space and Astronautical Science (ISAS), 
     Japan Aerospace Exploration Agency (JAXA), 3-1-1 Yoshinodai, Sagamihara, 229-8510, Japan; \myemail}
\altaffiltext{3}{Department of Physics and Astronomy, Rutgers University, 136 Frelinghuysen Road, Piscataway, NJ 
     08854-8019, USA; jph@physics.rutgers.edu}
\altaffiltext{4}{Center for Computational Astrophysics, Flatiron Institute, 162 Fifth Avenue, New York, NY 10010, USA}

\begin{abstract}
We report measurements of proper motion, radial velocity, and
elemental composition for 14 compact X-ray bright knots in Kepler's
supernova remnant (SNR) using archival \chandra\ data.  The highest
speed knots show both large proper motions ($\mu \sim$
0.11--0.14$^{\prime\prime}$ yr$^{-1}$) and high radial velocities ($v
\sim$ 8,700--10,020 km s$^{-1}$).  For these knots the estimated space
velocities (9,100 km s$^{-1}$ $\lesssim v_{\rm 3D} \lesssim$ 10,400 km
s$^{-1}$) are similar to the typical Si velocity seen in SN Ia near
maximum light. High speed ejecta knots appear only in specific
locations and are morphologically and kinematically distinct from the
rest of the ejecta.  The proper motions of five knots extrapolate back
over the age of Kepler's SNR to a consistent central position.  This
new kinematic center agrees well with previous determinations, but is
less subject to systematic errors and denotes a location about which
several prominent structures in the remnant display a high degree of
symmetry.  These five knots are expanding at close to the free
expansion rate (expansion indices of 0.75 $\lesssim m \lesssim$ 1.0),
which we argue indicates either that they were formed in the explosion
with a high density contrast (more than 100 times the ambient density)
or that they have propagated through relatively low density ($n_{\rm
  H} < 0.1$ cm$^{-3}$) regions in the ambient medium. X-ray spectral
analysis shows that the undecelerated knots have high Si and S
abundances, a lower Fe abundance and very low O abundance, pointing to
an origin in the partial Si-burning zone, which occurs in the outer
layer of the exploding white dwarf for SN Ia models.  Other knots show
slower speeds and expansion indices consistent with decelerated ejecta
knots or features in the ambient medium overrun by the forward shock.
Our new accurate location for the explosion site has well-defined
positional uncertainties allowing for a great reduction in the area
to be searched for faint surviving donor stars under non-traditional
single-degenerate SN Ia scenarios; because of the lack of bright stars
in the search area the traditional scenario remains ruled out.

\end{abstract}

\maketitle

\section{Introduction}\label{sec:intro}

Kepler's supernova (SN 1604) is one of the most well-studied young
supernova remnants (SNRs) in the Galaxy.  General consensus holds,
even without a light echo spectrum, that Kepler's SNR is a Type Ia SN
based largely on X-ray observations showing shocked ejecta with strong
silicon, sulfur, and iron emission and a near absence of oxygen
emission \citep[e.g.,][]{2007ApJ...668L.135R}. Multiple lines of
evidence going back decades
\citep[e.g.,][]{vdbk77,dennefeld82,whitelong83, hugheshelfand85} have
shown that Kepler's SNR is interacting with a dense (few particles per
cm$^{-3}$), strongly asymmetric, nitrogen-rich ambient medium.  More recent
work shows that localized regions in the SNR show prominent oxygen,
neon, and magnesium X-ray emission with nearly solar O/Fe abundance
ratios that indicate an association
\citep[e.g.,][]{2007ApJ...668L.135R,2013ApJ...764...63B,2015ApJ...808...49K}
with a dense circumstellar medium (CSM).  In these regions, infrared
observations reveal strong silicate features suggestive of the wind
from an O-rich asymptotic giant branch (AGB) star
\citep{2012ApJ...755....3W}.

\begin{figure*}[t]
 \begin{center}
  \includegraphics[trim=5mm 45mm 5mm 45mm,clip,width=14cm]{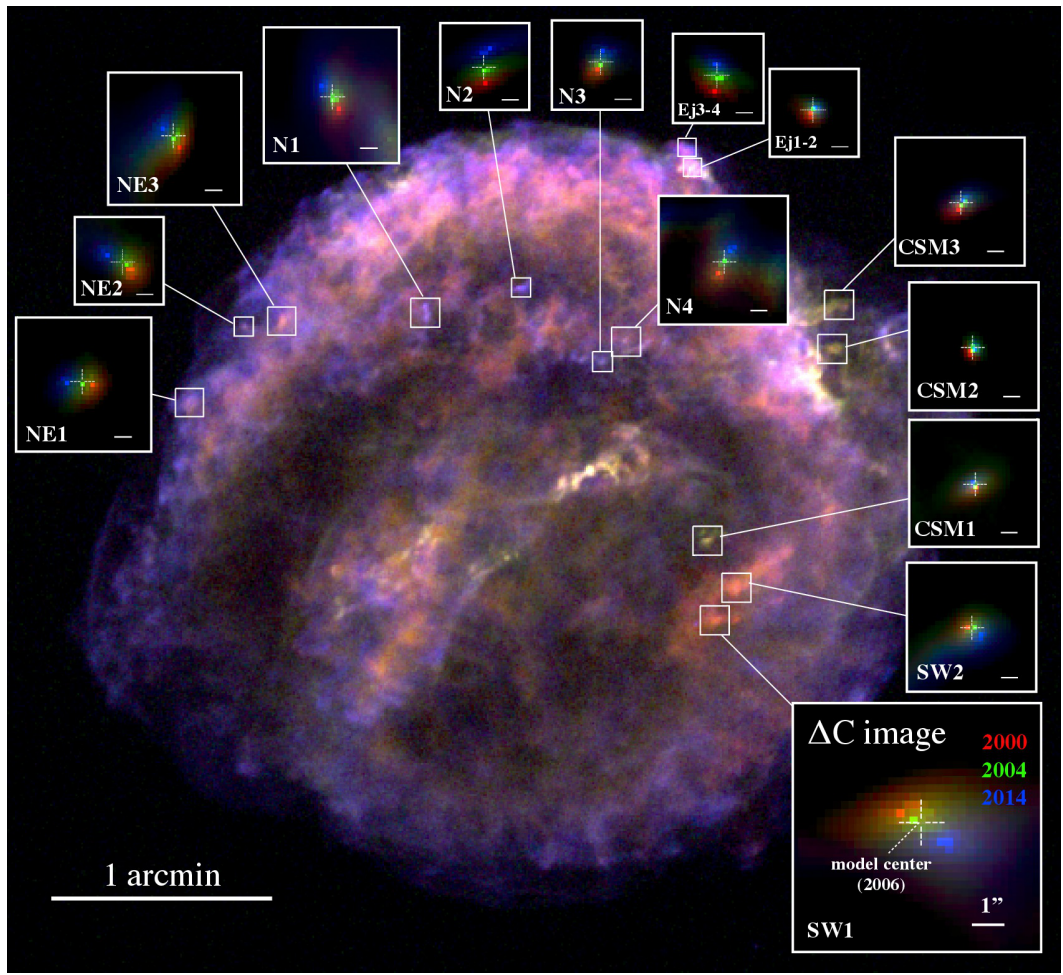}
 \end{center}
\caption{Three-color image of Kepler's SNR from the 2006 data set with
  red, green and blue images taken from the bands containing Fe
  L-shell emission (0.72-0.9 keV), O-Ly$\alpha$ emission (0.6--0.72
  keV), and Si-He$\alpha$ emission (1.78-1.93 keV), respectively. The
  image is binned by 0.246$^{\prime\prime}$ and has been smoothed with
  a Gaussian kernel with $\sigma = 0.492^{\prime\prime}$.  The
  intensity scale is square root. White boxes show the regions used
  for the proper motion analysis. Small frames around the figure show
  $\Delta C$ images from the image fits. The pure red, green and blue
  colors show fitting results of each knot's proper motion comparing
  the 2006 image to data from 2000, 2004 and 2014, respectively. The
  center position of each knot in the 2006 image is noted with the
  plus sign in each small frame, along with a 1$^{\prime\prime}$ scale
  bar in the lower right corner.  The brightest pixel in each color in
  the insert frames shows the minimum $C$ value for that epoch; the
  range of $C$ values plotted in each frame was adjusted to enhance
  visibility. We alert the reader to the different uses of color in
  this figure: the main panel uses color to show spectral variations
  with position in Kepler's SNR while the 14 insert panels use color
  to denote proper motions, showing fit results for the 3 different
  epochs with different colors. }
\label{fig:3color}
\end{figure*}

\citet{bandiera87} first connected the environmental characteristics
of Kepler's SNR, unusual for its location a few hundred pc above the
Galactic plane, with the possibility of a high-speed ($\sim$300 km
s$^{-1}$), mass-losing progenitor star. In his model, wind material is
compressed by the low density ambient medium forming a dense bow shock
in the direction of motion (toward the northwest).  This produces a
strong brightness gradient in X-ray and radio images and high density
knots in the resulting SNR (see, e.g., \citealt{borkowski+92} for an
early 2D hydro bow-shock model for the remnant). Bandiera's model is
likewise consistent with X-ray expansion measurements
\citep{2008ApJ...689..231V, 2008ApJ...689..225K} that show slower
expansion rates in the north compared with the rest of the remnant.
More recent multi-dimensional hydrodynamical models
\citep{velazquez+06, 2012A&A...537A.139C, 2013ApJ...764...63B,
  toledo-roy+14} have considered a single degenerate (SD) scenario for the
progenitor to Kepler's SNR with the mass loss coming from the donor
companion star to the white dwarf that exploded.  However, it is
important to note that no surviving red giant, AGB or post AGB donor
star has been found in the central region of Kepler's SNR
\citep{2014ApJ...782...27K}.

In this article, we report three-dimensional space velocities of
several X-ray knots in Kepler's SNR determined from both proper
motions and radial velocities using archival {\it Chandra}
observations.  Our results yield insights on the nature of the
explosion and the ambient medium around Kepler's SNR. In \S 2 we
present the observational results from image and spectral analysis of
the several knots, including measurements of elemental composition in
addition to velocity. The discussion section (\S3) studies the
kinematics of the knots; reports a new, accurate, kinematic center for
Kepler's SNR; assesses the implications of these results on the
evolutionary state and nature of the ambient medium; and reexamines
the question of a possible left-over companion star under the SD
scenario for the explosion. The final section summarizes the
article. Uncertainties are quoted at the 1$\sigma$ (68.3\%) confidence
level unless otherwise indicated; positions are given in equinox J2000
throughout.

\begin{figure*}[t]
 \begin{center}
  \includegraphics[trim=0mm 40mm 0mm 40mm,clip,width=14cm]{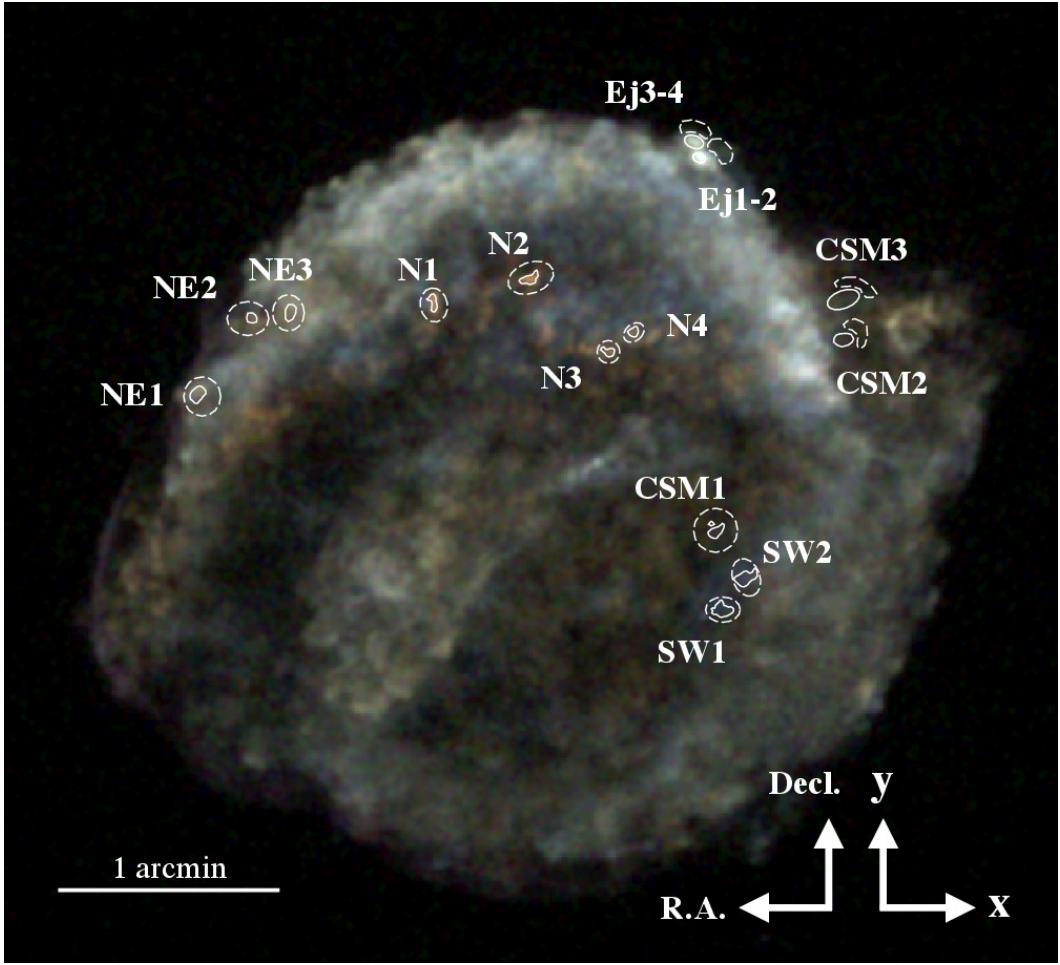}
 \end{center}
\caption{Three-color image showing the Doppler velocities of Kepler's
  SNR in the Si-He$\alpha$ line from the 2006 \chandra\ ACIS-S
  observation. The red, green and blue images come from the 1.78--1.83
  keV, 1.84--1.87 keV, and 1.88--1.93 keV bands. Each image is
  smoothed with a Gaussian kernel with $\sigma = 0.738^{\prime\prime}$
  Solid lines show the regions used for the spectral analysis and
  dashed curves show the background regions.  }
\label{fig:img-vel}
\end{figure*}

\section{Observational Results}

The {\it Chandra} Advanced CCD Imaging Spectrometer Spectroscopic-array
\citep[ACIS-S,][]{garmire+92,bautz+98} 
observed Kepler's SNR four times: in 2000
(PI: S.\ Holt), 2004 (PI: L.\ Rudnick), 2006 (PI: S.\ Reynolds) and
2014 (PI: K.\ Borkowski). The total net exposure time for each of
these observations is 48.8 ks, 46.2 ks, 741.0 and 139.1 ks,
respectively. The time differences with respect to the long
observation in 2006 are 5.96 yr (2000-2006), 1.64 yr (2004-2006) and
7.90 yr (2006-2014). We reprocessed all the level-1 event data,
applying all standard data reduction steps with CALDB version 4.7.2,
using a custom pipeline based on ``\verb"chandra_repro"'' in CIAO
version 4.8.

Serendipitous point sources were used to align the images.  Sources
were identified in each ObsID using the CIAO task \verb"wavdetect" and
position offsets were computed with \verb"wcs_match".  All ObsIDs were
matched to ObsID 6175, which was chosen as the reference because it
has the longest exposure time (159.1 ks). At least 4 and as many as 17
sources, depending on ObsID, were used for the alignment.  The source
positions showed mean shifts in R.A.\ and decl.\ of less than
0.35$^{\prime\prime}$. Using the shift values, we updated the aspect
solution of the event file using \verb"wcs_update". After corrections,
the average residuals in the point source positions relative to the
reference ObsID (\#6175) are $<$0.25$^{\prime\prime}$.

\subsection{Proper Motions of the Knots}

Figure \ref{fig:3color} shows a three-color image of Kepler's SNR from
the third epoch observation (in 2006) after image alignment,
highlighting emission from primarily O-Ly$\alpha$ (green), Fe L-shell
(red), and Si-He$\alpha$ (blue). Regions of CSM emission in Kepler's
SNR (e.g., CSM1--3 in the figure) tend to appear greenish in this
figure due to relatively more emission from O. Ejecta knots appear as
purple (both strong Si and Fe emission, e.g., N and NE knots) or
orange (weaker Si to Fe, e.g., SW knots). Some of the purplish colored
regions around the edge of the SNR contain strong nonthermal emission.

\begin{figure}[h]
 \begin{center}
  \includegraphics[width=8cm]{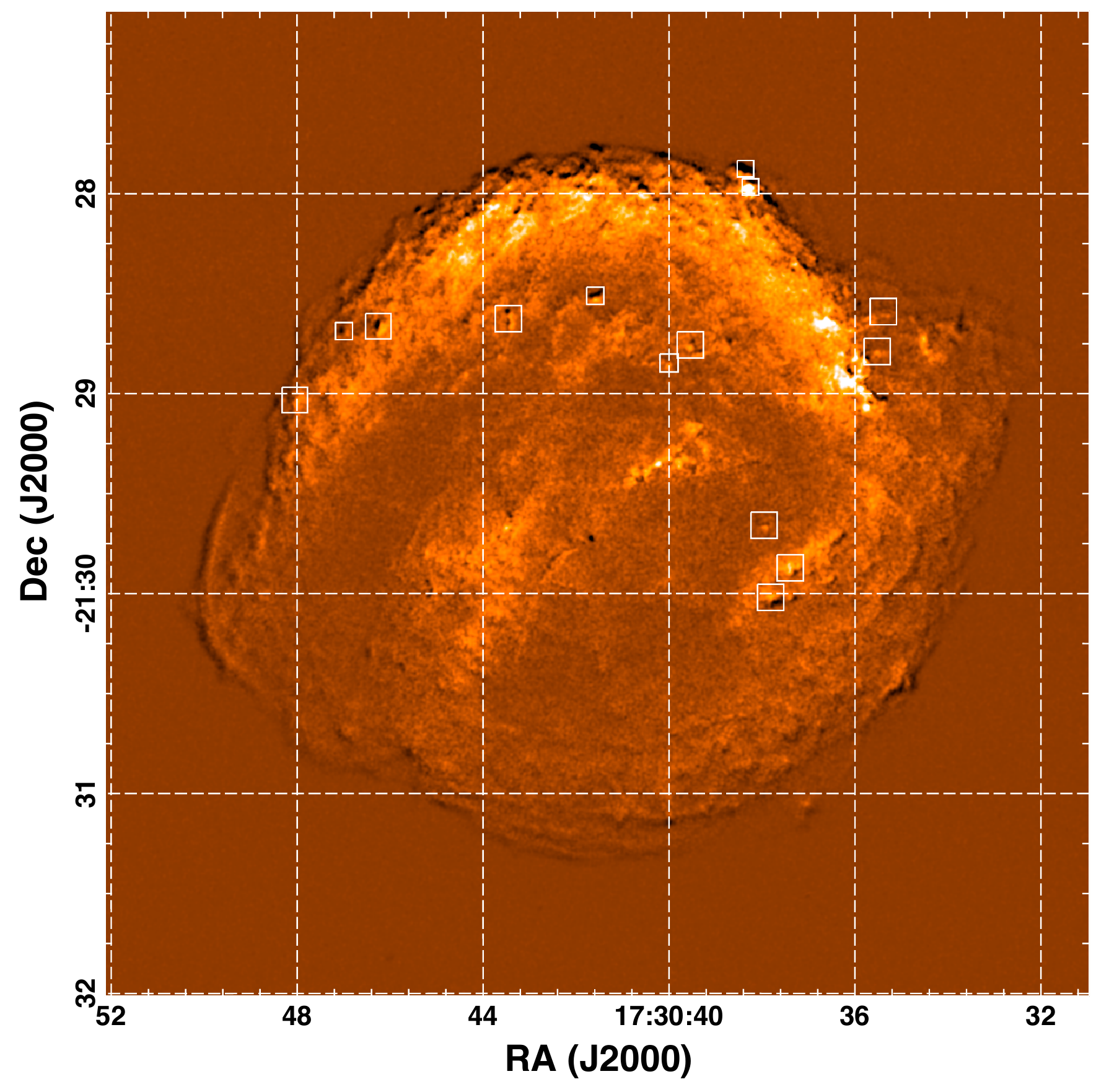}
 \end{center}
\caption{Difference image made by subtracting the two
  \chandra\ observations of Kepler's SNR taken in 2014 and 2000.
  White boxes show the regions used for the proper motion analysis.  }
\label{fig:imgdiff}
\end{figure}

Figure \ref{fig:img-vel} shows a Doppler velocity map of Kepler's SNR
made using three narrow energy bands in the Si-He$\alpha$ line. Our
recent work on Tycho's SNR \citep{2017ApJ...840..112S} demonstrated
the ability of the \chandra\ ACIS instrument to measure radial
velocities for relatively large and diffuse ejecta knots.  The radial
velocities of knots in Tycho's SNR show an obvious pattern indicative
of a spherically expanding shell---with the highest speeds through the
center and decreasing speeds towards the limb.
This pattern is not seen in Kepler's SNR where instead the highest
speeds (both red- and blue-shifted) appear as distinct knots lying in
chains that stretch east-west in specific locations largely across the
northern half of the remnant.  A number of these knots (the N, NE, and
SW sets) also showed large proper motions when comparing the images
taken in 2000 and 2014 (see Fig.~\ref{fig:imgdiff}).  We also
identified three CSM knots and the northwest ejecta knots (Ej1, 2, 3,
and 4, which we combined into two separate knots Ej1-2, Ej3-4) whose
proper motions were recently measured with the {\it Hubble Space
  Telescope} \citep[{\it HST};][]{2016ApJ...817...36S}. None of the N,
NE, and SW knots showed any evidence for optical emission in the {\it
  HST} images, while the CSM and Ej knots all did. We selected 14
knots for proper motion and radial velocity analyses.

To measure proper motions, we extracted image cut-outs about each knot
for the 4 epochs.  The image from the long observation in 2006 was
used as the fitting model for each knot and was shifted in
R.A.\ and decl.\ to obtain the 2D proper motion shifts. We employed the
C-statistic, a maximum likelihood statistic for Poisson distributed
data \citep{1979ApJ...228..939C},
\begin{equation}
 C = -2\Sigma_{i,j}(n_{i.j}~{\rm ln}~m_{i,j}-m_{i,j}-~{\rm ln}~n_{i.j}!)
\end{equation}
where $n_{i,j}$ are the counts in pixel ({\it i,j}) of the image in
each epoch, and $m_{i,j}$ are the model counts from the 2006 image
scaled by the relative number of total counts (over the 0.6--2.7 keV
band) from the entire SNR.  We estimate the errors on the proper
motion shifts using $\Delta C = C - C_{\rm min}$, which is similar to
$\chi^2$ \citep{1979ApJ...228..939C}. The fitting errors are
subdominant to the systematic errors in image alignment, which we
determine by fitting the positions of seven serendipitous point
sources using the same method. The systematic errors ($\sigma_{\rm
  x}$, $\sigma_{\rm y}$) are estimated to be (0.17$^{\prime\prime}$,
0.18$^{\prime\prime}$) for 2000, (0.26$^{\prime\prime}$,
0.20$^{\prime\prime}$) for 2004 and (0.21$^{\prime\prime}$,
0.32$^{\prime\prime}$) for 2014.

Images of $\Delta C$ for each knot are shown in small frames on Figure
\ref{fig:3color}, clearly indicating the high proper motion of the
ejecta knots, and Table \ref{tab:proper} presents numerical results.
We obtained acceptable fits for all knots (reduced $C$ =
0.95--2.04). For the NE, N and SW sets, we found large proper motions
($\sim$0.08--0.14$ ^{\prime\prime}$ yr$^{-1}$), comparable to values
from around the rim \citep[0.076--0.302$^{\prime\prime}$ yr$^{-1}$;
][]{2008ApJ...689..225K}. In contrast, the CSM knots have small proper
motions ($\lesssim$ 0.04$^{\prime\prime}$ yr$^{-1}$), consistent with
being either ejecta knots significantly decelerated by interaction
with the CSM or features in the ambient medium overrun by the forward
shock.  Our X-ray proper motion of the Ej3-4 knot is consistent with
the H$\alpha$ result \citep[$\sim$0.08--0.112$^{\prime\prime}$
  yr$^{-1}$; ][]{2016ApJ...817...36S}, but the proper motion of knot
Ej1-2 is smaller by about a factor of two in the X-rays than in
H$\alpha$ (0.069--0.083$^{\prime\prime}$ yr$^{-1}$).  Knot Ej3-4 is
detached from the main shell of X-ray emitting ejecta (and relatively
distant from any slow moving optical knots); the agreement between the
\chandra\ and {\it HST} proper motions indicates that the X-ray knot
is driving the arc-shaped H$\alpha$ shock here.  X-ray knot Ej1-2,
however, is closer to the (slower moving) main ejecta shell and is
partially superposed on bright optical radiative knots that appear to
be slowly moving.  We suspect that the discrepancy between the motion
of the H$\alpha$ shock and knot Ej1-2 is due to some contamination of
the X-ray knot by slower moving ejecta from the main shell or X-ray
emission associated with the optical radiative knots.

\begin{table*}[t]
\caption{Proper Motions and Radial Velocities of  Knots in Kepler's SNR}\label{tab:proper}
\begin{center}
\begin{tabular}{l|cccc|cccccccc}
\hline
		&																							\multicolumn{4}{|c|}{\bf Imaging Analysis}														&	\multicolumn{2}{|c}{\bf Spectral Analysis}					\\
		&		center of the model frame	(epoch 2006)														&		mean sift: $\Delta$x, $\Delta$y			&	proper motion		&	angle				&	radial velocity 				&	$\chi^2_{\nu}$ (d.o.f)	\\
id		&		 R.A., Decl.\ (J2000)																		&			(arcsec yr$^{-1}$)					&	(arcsec yr$^{-1}$)	&	(degree)			&		(km s$^{-1}$)				&							\\\hline
NE1		& 17$^{\rm h}$30$^{\rm m}$48$^{\rm s}$.022, $-21^\circ$29$^\prime$01$^{\prime\prime}$.87	&	$-0.112\pm$0.020, $+0.018\pm$0.019			&	0.112$\pm$0.020		&	81$\pm$10			&	$+$3170$^{+50}_{-200}$			&		1.05 (320)			\\
NE2		& 17$^{\rm h}$30$^{\rm m}$46$^{\rm s}$.963, $-21^\circ$28$^\prime$41$^{\prime\prime}$.21	&	$-0.099\pm$0.021, $+0.064\pm$0.020			&	0.117$\pm$0.021		&	57$\pm$10			&	$+$6540$^{+890}_{-570}$			&		0.92 (166)			\\
NE3		& 17$^{\rm h}$30$^{\rm m}$46$^{\rm s}$.224, $-21^\circ$28$^\prime$39$^{\prime\prime}$.74	&	$-0.064\pm$0.020, $+0.081\pm$0.020			&	0.103$\pm$0.020		&	38$\pm$11			&	$+$2970$^{+210}_{-220}$			&		1.26 (297)			\\
N1		& 17$^{\rm h}$30$^{\rm m}$43$^{\rm s}$.428, $-21^\circ$28$^\prime$37$^{\prime\prime}$.47	&	$-0.063\pm$0.020, $+0.088\pm$0.020			&	0.108$\pm$0.020		&	36$\pm$11			&	$+$8700$^{+650}_{-470}$			&		0.96 (258)			\\
N2		& 17$^{\rm h}$30$^{\rm m}$41$^{\rm s}$.561, $-21^\circ$28$^\prime$30$^{\prime\prime}$.59	&	$+0.008\pm$0.020, $+0.141\pm$0.019			&	0.141$\pm$0.019		&	357$\pm$8			&	$+$9110$^{+30}_{-110}$			&		1.32 (259)			\\
N3		& 17$^{\rm h}$30$^{\rm m}$39$^{\rm s}$.974, $-21^\circ$28$^\prime$50$^{\prime\prime}$.75	&	$+0.029\pm$0.020, $+0.076\pm$0.020			&	0.081$\pm$0.020		&	339$\pm$14			&	$+$5880$^{+690}_{-1750}$		&		0.82 (165)			\\
N4		& 17$^{\rm h}$30$^{\rm m}$39$^{\rm s}$.516, $-21^\circ$28$^\prime$45$^{\prime\prime}$.34	&	$+0.048\pm$0.021, $+0.099\pm$0.020			&	0.110$\pm$0.020		&	334$\pm$11			&	$+$10020$^{+1270}_{-440}$		&		0.86 (192)			\\
SW1		& 17$^{\rm h}$30$^{\rm m}$37$^{\rm s}$.788, $-21^\circ$30$^\prime$01$^{\prime\prime}$.09	&	$+0.101\pm$0.021, $-0.060\pm$0.020			&	0.118$\pm$0.020		&	239$\pm$10			&	$-$5590$^{+340}_{-260}$			&		1.01 (253)			\\
SW2		& 17$^{\rm h}$30$^{\rm m}$37$^{\rm s}$.366, $-21^\circ$29$^\prime$52$^{\prime\prime}$.23	&	$+0.068\pm$0.020, $-0.041\pm$0.020			&	0.079$\pm$0.020		&	239$\pm$14			&	$-$8000$^{+500}_{-150}$			&		1.24 (297)			\\
CSM1	& 17$^{\rm h}$30$^{\rm m}$37$^{\rm s}$.930, $-21^\circ$29$^\prime$39$^{\prime\prime}$.44	&	$-0.005\pm$0.020, $+0.022\pm$0.019			&	0.023$\pm$0.019		&	14$\pm$49			&	$+$740$^{+980}_{-490}$			&		1.14 (191)			\\
CSM2	& 17$^{\rm h}$30$^{\rm m}$35$^{\rm s}$.498, $-21^\circ$28$^\prime$47$^{\prime\prime}$.31	&	$+0.028\pm$0.020, $+0.023\pm$0.019			&	0.037$\pm$0.019		&	309$\pm$30			&	$-$2300$^{+240}_{-280}$			&		1.19 (227)			\\
CSM3	& 17$^{\rm h}$30$^{\rm m}$35$^{\rm s}$.366, $-21^\circ$28$^\prime$35$^{\prime\prime}$.31	&	$+0.021\pm$0.020, $+0.039\pm$0.019			&	0.044$\pm$0.019		&	332$\pm$26			&	$+$574$^{+7}_{-90}$				&		1.26 (309)			\\
Ej1-2	& 17$^{\rm h}$30$^{\rm m}$38$^{\rm s}$.222, $-21^\circ$27$^\prime$57$^{\prime\prime}$.92	&	$+0.028\pm$0.020, $+0.026\pm$0.019			&	0.038$\pm$0.019		&	313$\pm$29			&	$+$244$^{+46}_{-10}$			&		1.42 (445)			\\
Ej3-4	& 17$^{\rm h}$30$^{\rm m}$38$^{\rm s}$.328, $-21^\circ$27$^\prime$52$^{\prime\prime}$.51	&	$+0.010\pm$0.020, $+0.112\pm$0.019			&	0.112$\pm$0.019		&	355$\pm$10			&	$+$351$^{+19}_{-22}$			&		1.69 (395)			\\
\hline
\end{tabular}
\end{center}
\end{table*}

\begin{figure}[h]
 \begin{center}
  \includegraphics[width=8cm]{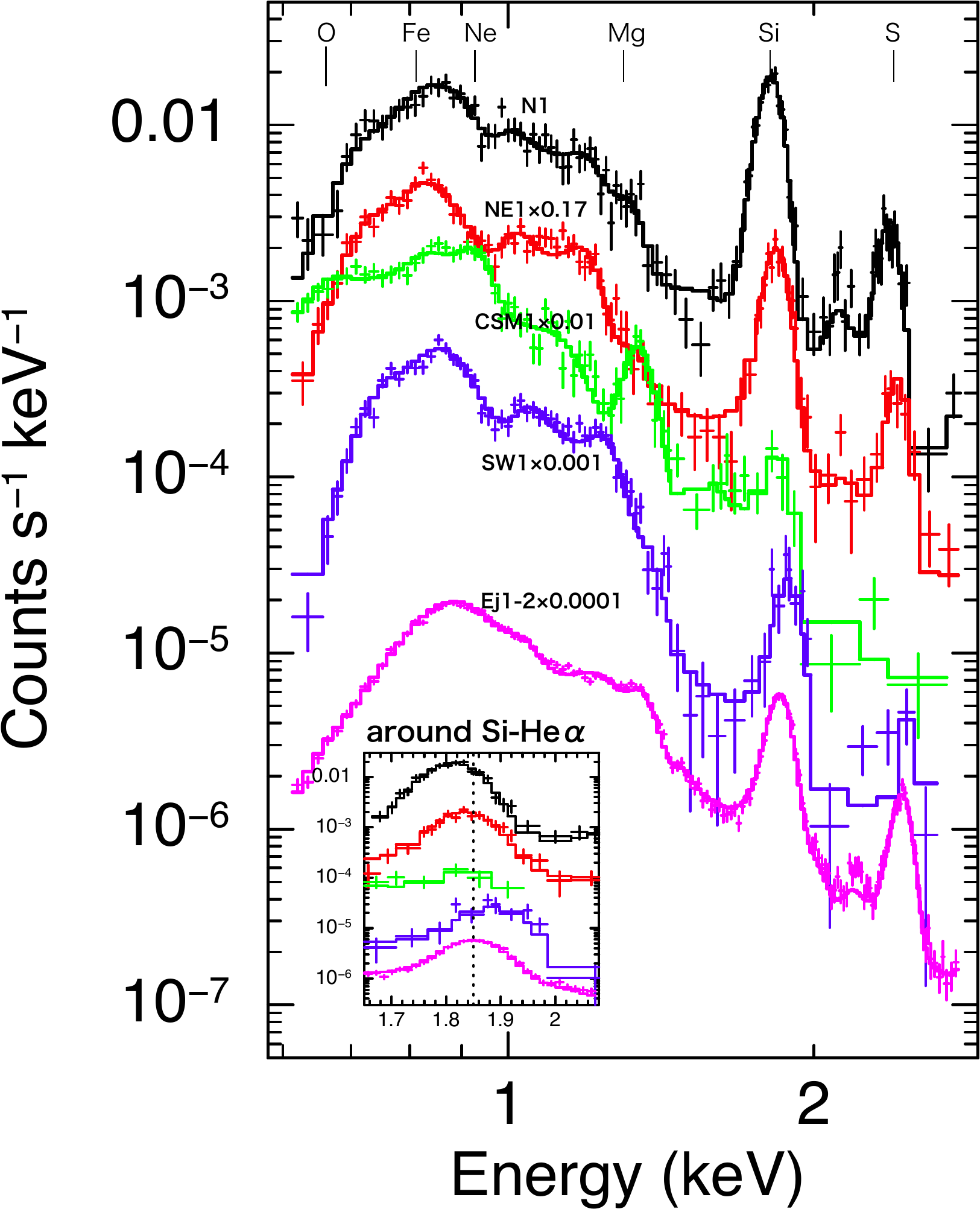}
 \end{center}
\caption{Observed spectra and best fitting models for several knots in
  Kepler's SNR (black: N1, red: NE1, green: CSM1, blue: SW1, magenta:
  Ej1-2). Although we only show the spectra from the 2006 data set
  here, spectra from all four epochs were used to constrain the model
  in the joint fit.  The inset plots the spectra in the vicinity of the
  Si line, with an expanded energy scale, in order to illustrate the
  Doppler shifts. }
\label{fig:spec}
\end{figure}

\begin{figure*}[t]
 \begin{center}
  \includegraphics[width=18cm]{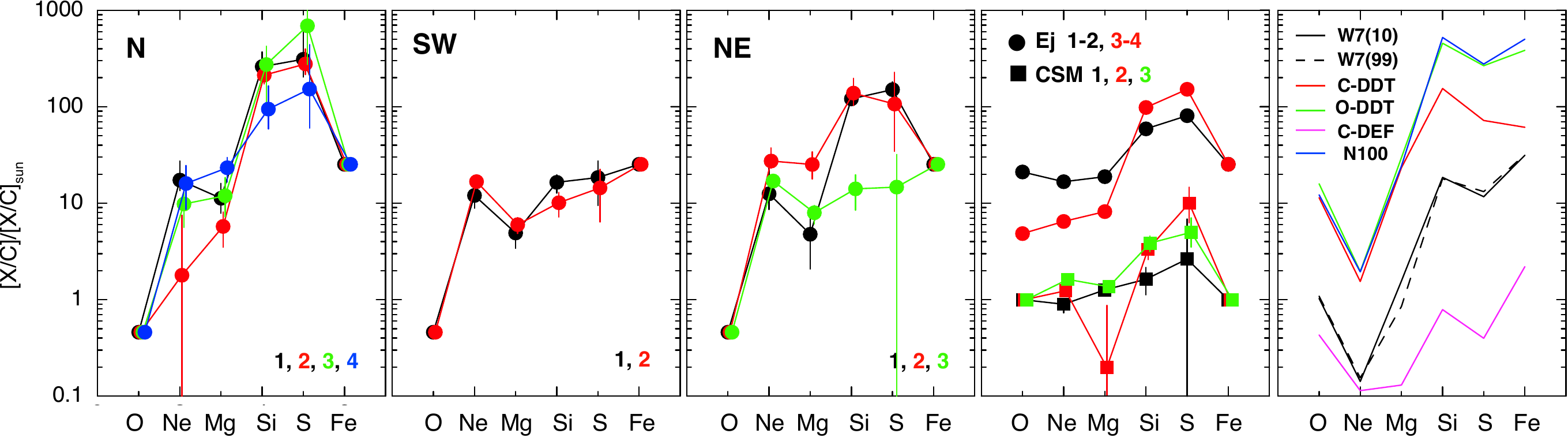}
 \end{center}
\caption{Plots of the relative abundances of various elemental species
  with respect to carbon relative to solar ratios for the several sets
  of knots studied here. The rightmost panel shows, for comparison,
  the integrated yields from a variety of SN Ia explosion models
  including W7(99): W7 from \citet{iwamoto+99}; W7(10): W7 from
  \citet{maeda+10}; C-DEF: spherically symmetric pure-deflagration,
  C-DDT: delayed detonation after a spherical deflagration, and O-DDT:
  delayed detonation after an extremely offset deflagration all from
  \citet{maeda+10}; and N100: delayed detonation after a deflagration
  initiated at 100 ignition spots from \citet{seitenzahl+13}.}
\label{fig:abun}
\end{figure*}

\subsection{X-ray Spectroscopy of the Knots}

We extracted spectra in each epoch (2000, 2004, 2006 and 2014) from
the knot and background regions defined in Figure \ref{fig:img-vel},
accounting for position shifts due to the proper motion.
The spectra were fitted in the 0.6--2.8 keV band using an absorbed
\verb"vvnei + power-law" model in {\tt XSPEC} 12.9.0 (AtomDB v3.0.3).  An
additional Gaussian model was included to account for a feature at
$\sim$1.2 keV from missing Fe-L lines in the atomic database
\citep{2000ApJ...530..387B,2001A&A...365L.329A}. 
Among the four spectra from the different epochs for each knot, fitted
model parameters (temperature, ionization age, abundances, and radial
Doppler velocity) were linked.
Our spectral fits explicitly allow
the ionization timescale, temperature, and redshift to be free
parameters, so line centroid variations due to changes in the
thermodynamic state are explicitly included in our fits, the derived
values, and the uncertainty on the fitted redshift.
To reduce the
complexity of our fits, we fixed a number of our model parameters to
the best-fit values determined by \cite{2015ApJ...808...49K}.
Specifically we fixed the column density to the value $N_H = 6.4
\times 10^{21}$ cm$^{-2}$ using the abundance table from
\citet{2000ApJ...542..914W} and the photon index of the power-law
model to 2.64, allowing the normalization parameter to be free.  We
assumed no H and He in the shocked SN Ia ejecta (for the N, NE, SE and
Ej knots), and fitted the Ne, Mg, Si and S abundances as free
parameters. We fixed the abundances of [O/C]/[O/C]$_{\odot}$,
[Ar/C]/[Ar/C]$_{\odot}$, [Ca/C]/[Ca/C]$_{\odot}$ and
[Fe/C]/[Fe/C]$_{\odot}$ to be 0.46, 37, 67.41 and 25.28, respectively.
For the Ej knots, we had to thaw the O abundance to obtain good fits.
Our fits for these knots also showed more neon and magnesium emission
compared to the other ejecta knots, implying that the Ej knots show a
mix of both ejecta and CSM components.  For the CSM knots, we used
only an absorbed \verb"vvnei" model.  Here we include hydrogen and
helium in the plasma, and fixed the abundances of He, C, O, Ar, Ca and
Fe to the solar values.  The nitrogen abundance is fixed to
[N/H]/[N/H]$_{\odot} = 3.5$ as expected for the N-rich CSM of Kepler's
SNR, while the Ne, Mg, Si and S abundances are allowed to be free
parameters. Figure
\ref{fig:spec} shows example spectra and best fitting models; the
small insert figure shows the clear effect of the Doppler shifting on
the Si-line of the knot spectra.

The first four panels of figure~\ref{fig:abun} plot the abundances of
the various knots from our spectral fits.  The abundances of all the N
knots plus NE1 and NE2 are quite similar and show strong enhancements
of Si and S compared to Fe as well as the lighter elements.  This is
also the case for the Ej knots, but with less contrast between Si/S
and the other species. The SW knots plus NE3 appear to follow a
different abundance pattern, while the CSM knot abundances are close
to the solar ratios for Ne and Mg, but show some enhancement (by
factors of a few) in Si and S (albeit with large uncertainties).  The
last panel of this figure shows integrated yields from a variety of
published SN Ia explosion models (see the figure caption for details
and citations) that indicate that the ejecta knot abundances we find
are broadly consistent with theoretical expectations.  However, it is
highly unlikely that the knots should contain material with the
spatially integrated yields, but rather they should reflect the
composition of the location in the exploding white dwarf where they
were formed. Indeed the high abundances of Si and S relative to Fe
(especially for NE1 and NE2 and the N knots) indicate that these knots
formed at the outer, partially burned layer of the exploding white
dwarf.  The presence of Fe and low O abundance further restrict their
origin to the partial Si-burning regime (e.g., mass coordinate range
of 0.7--0.9 $M_\odot$ in W7, \citealt{nomoto+84}).  A future study
will use the fitted abundances in more detail to better identify where
these knots formed in the explosion.

Our fits are able to accurately represent the various knot spectra
across the range of significant compositional differences we
obtain. We also find that the best-fitting electron temperatures
($kT_{\rm e}$) and ionization ages ($n_{\rm e}t$) differ from knot to
knot ($kT_{\rm e} \sim 0.2 - 1$ keV, $n_{\rm
  e}t\sim7\times10^{9}-5\times10^{11}$ cm$^{3}$ s).  In some cases
these are correlated in interesting ways.  For example the O abundance
and ionization age in knot Ej1-2 are larger than those in the adjacent
Ej3-4 knot: [O/C]/[O/C]$_{\odot} = 19 -23$ and $n_{\rm e}t = (1.5
-1.6)\times10^{11}$ for Ej1-2, [O/C]/[O/C]$_{\odot} = 4-5$ and $n_{\rm
  e}t = (5.9-6.9)\times10^{10}$ for Ej3-4.  These spectral results
argue for a significant CSM interaction for knot Ej1-2, in agreement
with the argument put forward above to explain the proper motion
differences.

Radial velocity determination can be sensitive to the background
subtraction. To test this, we replaced the individual local background
regions with an annular ($r =2.4^{\prime}$--$3.5^{\prime}$) blank sky
region surrounding the remnant.  The spectral fits were as good, but
the precise values of the measured speeds differed, on average, by
$\sim$1,500 km s$^{-1}$ with the local background results being
higher than the blank sky background in nearly all regions (exceptions
were the CSM and Ej regions). This trend is expected because using
local background regions removes contaminating emission with a
different velocity (from, e.g., the other hemisphere of the remnant)
projected across the knot spectral extraction region. Such
contamination tends to reduce a knot's observed speed compared to its
actual speed.

The numerical accuracy of our velocity measurements with the ACIS-S
detector is limited by ACIS gain calibration
uncertainties\footnote{See http://web.mit.edu/iachec/ for
the current calibration status}.  For example,
\cite{2017ApJ...840..112S} showed a discrepancy in the radial velocity
measurements of $\sim$500--2,000 km s$^{-1}$ between the ACIS-S and
ACIS-I detectors for a set of knots similar to those we study here,
which we argued was likely a result of uncertain gain
calibration. Still, the large velocities we measure for most knots
($>$ 5,000 km s$^{-1}$) remain significant even given the level of
systematic uncertainty due to instrumental effects mentioned here and
background subtraction discussed in the previous paragraph.

The two rightmost columns of Table \ref{tab:proper} summarize the
radial velocity fits.  The N and SW knots have high radial speeds
(5,590 km s$^{-1}$ $< v <$ 10,020 km s$^{-1}$) that are $\sim$2--3
times higher than the ejecta knot speeds quoted by
\citet{2016ApJ...817...36S} from {\it HST} proper motions: $\sim$
1,600--3,000 km s$^{-1}$.  On the other hand, the CSM and Ej knots
show relatively low speeds ($<$ 2,300 km s$^{-1}$). 

\begin{figure*}[t]
 \begin{center}
  \includegraphics[width=18cm]{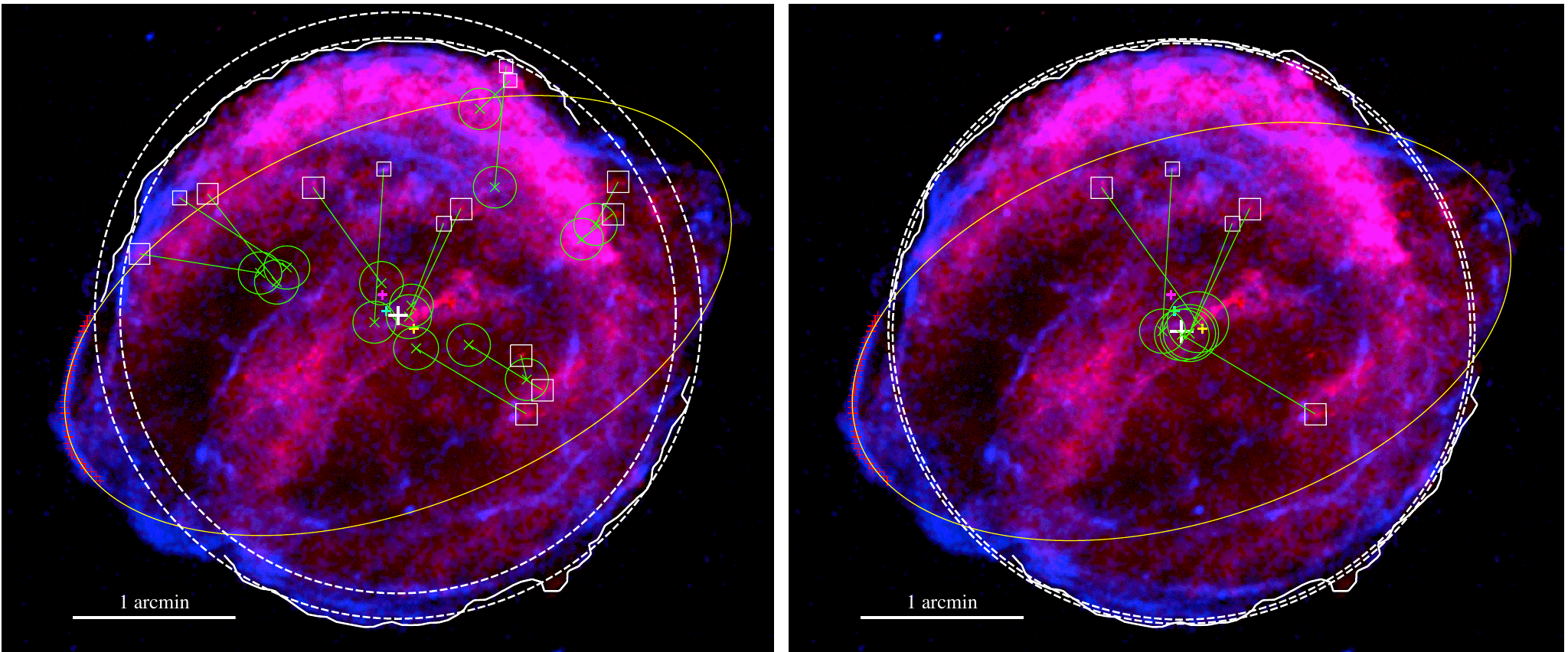}
 \end{center}
 \caption{Two-color image of the 0.6--2.7 keV band (red) and the
   4.2--6 keV band (blue).  {\it Left:} Extrapolation of the measured
   proper motion vector for each knot back to the explosion date (1604
   Oct 9), assuming purely undecelerated motion. The green lines,
   crosses and circles show the estimated distance moved, best-fit
   extrapolated original position and 1 $\sigma$ uncertainty. Five
   knots (N1, N2, N3, N4 and SW1) extrapolate back to a consistent
   position which we note by the solid white cross symbol (whose size
   denotes the 1 $\sigma$ uncertainty).  The other cross symbols
   denote expansion centers estimated by others: cyan
   \citep[radio:][]{1984ApJ...287..295M}, magenta and yellow
   \citep[X-ray:][]{2008ApJ...689..225K,2008ApJ...689..231V}. Two
   dashed circles, centered on the position we determine here, match
   the northern ($r = 1.71^{\prime}$) and southern ($r =
   1.87^{\prime}$) extent of Kepler's SNR.  The yellow ellipse is
   centered on the kinematic center, while the axis lengths and
   orientation are matched to the nonthermal filament on the eastern
   rim (highlighted with red crosses).  {\it Right:} In this panel, we
   use only the five knots with the highest expansion indices to
   extrapolate back to the explosion center including the effects of
   deceleration using the measured expansion index.  The agreement
   between the individual knots is greatly improved.  The dynamical
   center here is shifted by about 6$^{\prime\prime}$ south of the
   center shown in the left panel. }
\label{fig:cnt}
\end{figure*}

\section{Discussion}\label{discussion}

\subsection{Undecelerated Ejecta Knots and the Kinematic Center of Kepler's SNR}

\begin{figure}[h]
 \begin{center}
  \includegraphics[width=8cm]{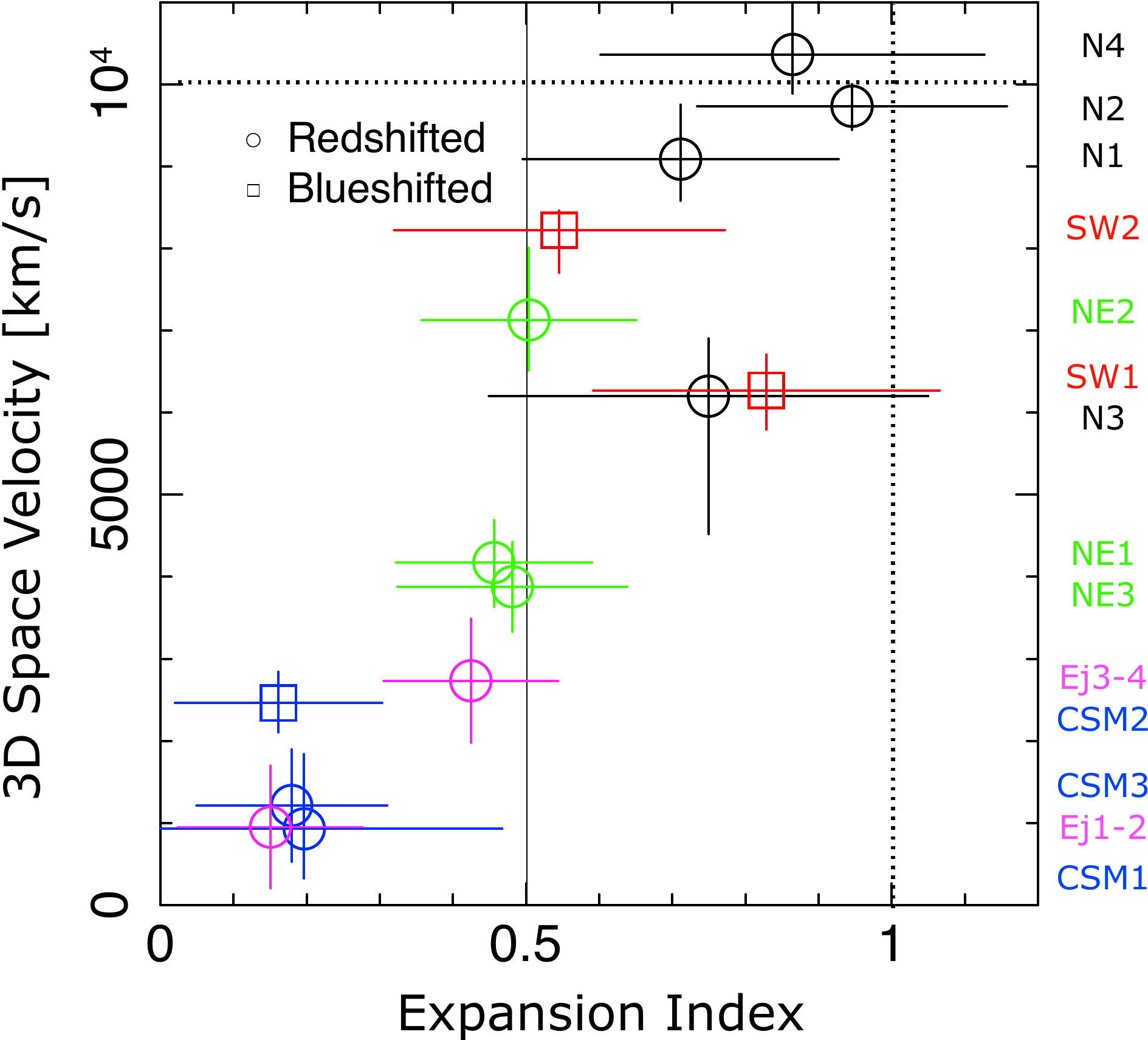}
 \end{center}
 \caption{Scatter plot between 3D space velocity and expansion index
   for the 14 knots identified in Fig.~\ref{fig:3color}.  The space
   velocity is the root-sum-square combination of the radial velocity
   and the proper motion assuming a distance of 5 kpc to Kepler's
   SNR. The plotted uncertainties are at the 90\% confidence
   level. The different knots are indicated with different colored
   symbols and are labeled along the right side. Circles (boxes)
   indicate redshifted (blueshifted) knots.  The solid vertical line
   shows the average expansion index for the remnant \citep[$r \propto
     t^{0.5}$:][]{2008ApJ...689..231V}.}
\label{fig:exp}
\end{figure}

Given the known age of Kepler's SNR ($401.7$ yr at the mean time of
the third epoch observation) we can use our proper motion vectors to
extrapolate the position of each knot back to its location at the time
of explosion. Initially, we assume that each knot moves without
any deceleration. Figure \ref{fig:cnt} (left panel) shows the 2006
locations (small boxes), the distance traveled (green lines) and the
initial locations in 1604 (green circles) of each of the knots.
Five knots extrapolate back to a consistent position (for values see
Table \ref{tab:center}), which we identify as the kinematic center of
the explosion.
This result agrees well with other estimates for the
explosion center, but, since it relies on knots that are nearly
undecelerated, it is much less sensitive to systematic errors due to
the spatial variation of the expansion rate across Kepler's SNR. We
use our kinematic center to determine the expansion index ($m$ in the
relation $r\propto t^m$) as $m = \mu \times t / r$, which depends only
on each knot's measured proper motion ($\mu$), its distance from the
expansion center ($r$) and the remnant's age ($t$).  The expansion
indices (see Fig.~\ref{fig:exp}) range from low values, $m < 0.25$,
indicating significant deceleration to high values, $m > 0.75$,
indicating little to no deceleration.

For a power-law evolution of radius with time as we use here, it is
possible to estimate the effects of deceleration on the distance
traveled by the knot and thereby obtain a more accurate estimate for
the kinematic center. We assume that the expansion index is constant
with time, which is only a good approximation for knots with high
expansion indices; so we restrict our analysis to the same five knots
introduced in the preceding paragraph.  We integrate the time
evolution of velocity $v = v_f (t/t_f)^{m-1}$ over the age of the SNR
($t_f$) to obtain the simple result for the distance traveled: $r=v_f
t_f / m$.  We start with the results from the previous paragraph,
which yielded a kinematic center and an estimate of $m$ for each knot.
The distance each knot has moved is now redetermined assuming
decelerated motion according to its $m$ value (although $m$ was
restricted to values of 1 or less) and the equation introduced a few
lines above.  Averaging these values leads to a new estimate for the
kinematic center.  This process was iterated until the individual $m$
values and the location of the kinematic center converged, which took
about 30 iterations. The kinematic center shifts slightly south from the
case of undecelerated motion above (see Table~\ref{tab:center}), but
the difference between the two estimates is not highly significant,
only $\sim$1 $\sigma$ ($\sim6^{\prime\prime}$).  Note how the $m$ values 
have changed only slightly as well.

\begin{table}[t]
\caption{Kinematic center of Kepler's SNR from high speed knots}\label{tab:center}
\begin{center}
\begin{tabular}{ccc}
\hline
& Undecelerated  & Decelerated  \\
Parameter & ($r=v_f t_f$) & ($r=v_f t_f/m$) \\ \hline
\multicolumn{3}{c}{Kinematic center}\\
R.A.  & 17$^{\rm h}$30$^{\rm m}$41$^{\rm s}$.189 & 17$^{\rm h}$30$^{\rm m}$41$^{\rm s}$.321 \\
Decl. & $-21^\circ$29$^\prime$24$^{\prime\prime}$.63 & $-21^\circ$29$^\prime$30$^{\prime\prime}$.51 \\
$\sigma_{\rm RA}$,  $\sigma_{\rm Dec}$  &  $\pm3.6^{\prime\prime}$, $\pm3.5^{\prime\prime}$  & $\pm4.4^{\prime\prime}$, $\pm4.3^{\prime\prime}$  \\\hline
\multicolumn{3}{c}{Expansion indices and offsets ($^{\prime\prime}$) from kinematic center}\\
 &  $m$, x, y & $m$, x, y \\
Knot N1         & 0.76, $-$12.4, $+$24.2                     & 0.71, $+$12.3, $+$7.0\\
Knot N2         & 1.04, $-$17.6, $\phantom{0}-$5.1           & 0.95, $-$14.3, $+$0.2\\
Knot N3         & 0.87, $+$10.1, $\phantom{0}+$7.5           & 0.75, $\phantom{0}+$5.8, $-$1.2\\
Knot N4         & 0.97, $\phantom{0}+$8.2, $\phantom{0}-$0.9 & 0.86, $\phantom{0}+$4.9, $-$1.6\\
Knot SW1        & 0.80, $+$13.5, $-$24.7                     & 0.83, $\phantom{0}+$0.2, $-$2.5\\
RMS x,y offsets    & 12.8, 16.0 & 9.1, 3.4\\ \hline
\multicolumn{3}{c}{Overall radius of Kepler's SNR from kinematic center}\\
Radius to N ($^\prime$)   & $1.71\pm0.04$ & $1.77\pm0.03$\\
Radius to S ($^\prime$)   & $1.87\pm0.04$ & $1.80\pm0.04$\\
\hline
\end{tabular}
\end{center}
\end{table}

Combining the proper motions and radial velocities allows us to
determine the 3-dimensional (3D) space velocities of these X-ray
knots.  The distance to Kepler's SNR is not well known with estimates
ranging from $\sim$4.0 kpc to $>$ 7 kpc; here we use a value of 5 kpc
which is consistent with \ion{H}{1} absorption measurements
\citep{1999AJ....118..926R} and recent optical proper motion
measurements of Balmer shocks \citep{2016ApJ...817...36S}.  Figure
\ref{fig:exp} presents the scatter plot of 3D space velocity versus
expansion index.  There is a clear trend for low space velocity knots
to have small expansion indices, while the high space velocity knots
tend to have large expansion indices. It is notable that the three
knots with the highest space velocities also have large expansion
indices; for these specific knots we determine space velocities of
9,100 km s$^{-1}$ $\lesssim v_{\rm 3D} \lesssim$ 10,400 km s$^{-1}$
and expansion indices of 0.75 $\lesssim m \lesssim$ 1.0.  Thus not
only are these knots expanding at nearly the free expansion rate, they
are moving with space velocities that are comparable to the expansion
speed of Si ejecta in SN Ia near maximum brightness
\citep[10,000--12,000 km s$^{-1}$, see][and references
  within]{1997ARA&A..35..309F}.

We can also estimate the 3D radial locations of the knots using each
knot's current projected distance from the kinematic center and the
angle defined by the radial and transverse (proper motion) speeds.
The angle obviously depends on the remnant's distance. As a reference
for comparison, we use the maximum projected extent of Kepler's SNR
from the kinematic center, which is $\sim$2.3$^\prime$ (3.35 pc for a
distance of 5 kpc); this occurs at the northwest protuberance. A knot
moving with a constant speed of 10,000 km s$^{-1}$ would reach a
radius of 4.1 pc in the lifetime of Kepler's SNR, and this radius
would be equal to the remnant's maximum projected extent for a
distance of 6.1 kpc.  For the fastest moving knots (N1, N2, and N4),
the estimated 3D radial locations are 1.57, 1.26, and 1.48 times the
maximum projected extent assuming a distance of 5 kpc. Values are
larger than the simple example given because these knots have suffered
some deceleration.  For a distance of 7 kpc, the radial locations are
1.16, 0.95, and 1.09, respectively, times the maximum projected
extent.  These considerations tend to favor distances to Kepler's SNR
on the larger range of those reported.

There are interesting relationships between the elemental composition of
the knots and their space velocities and inferred extent of
deceleration. The N series of knots show both high speed and low
deceleration ($m>0.5$), along with a clear ejecta-dominant composition
showing high Si and S abundances (Fig.~\ref{fig:abun}).  Although the
knots NE1 and NE2 show similar abundances, they appear to have
been decelerated more ($m \approx 0.5$) and are currently moving more
slowly.  The SW knots are kinematically similar to the N
knots, but are noticeably different in composition.
On the other hand, the CSM and Ej knots have low space velocities
($v_{\rm 3D} <$ 3,000 km s$^{-1}$) together with small expansion
indices ($m <$ 0.5) and have relatively less contrast between their
light (O, Ne, Mg) and heavy (Si, S, Fe) element abundances. Based on
the abundance patterns, the CSM knots appear to be dominated by
the ambient medium while the Ej knots are more dominated by ejecta.

\subsection{Global Evolution of Kepler's SNR}

The mean expansion index of Kepler's SNR from published studies using
high resolution \chandra\ images is $m\sim0.5$ with evidence for
higher rates of expansion in the south compared to the north
\citep[e.g.,][]{2008ApJ...689..225K,2008ApJ...689..231V}.  The dashed
circles plotted in Fig.~\ref{fig:cnt} are centered on our new
kinematic center for either undecelerated (left panel) or decelerated
(right panel) motion and were chosen to match the northern and
southern extents of the SNR.  The north/south radii (numerical values
given in Table~\ref{tab:center}) differ by $9^{+5}_{-4}$ \% for
undecelerated motion (left) and $2\pm4$ \% for decelerated motion
(right), where the uncertainties were estimated from the standard
deviation of the radial scatter of the plotted contour about the
estimated best-fit circle.

\cite{2012A&A...537A.139C} calculated models for Kepler's SNR assuming
the progenitor system was a symbiotic binary (a white dwarf and a 4-5
$M_\odot$ AGB star) moving toward the northwest with a velocity of 250
km s$^{-1}$.  As in the model of \citet{bandiera87}, this produces an
asymmetric wind around the progenitor with the densest regions at the
stagnation point (i.e., where the momentum of the wind and ambient
medium equilibrate) ahead of the star in the direction of motion.  In
their model A, which provides a decent description of Kepler's SNR,
the forward-shock interaction with the wind bubble begins $\sim$300 yr
after the explosion when the forward shock was at a radius of $\sim$2.7
pc.  This encounter has a strong, immediate, effect on the expansion
index in the direction toward the wind-stagnation point where the
dense shell material causes the forward shock to decelerate quickly. With
time, the radial asymmetry of the remnant also grows, but the decrease
of the expansion index is more immediate and dramatic.

As seen in Fig.~7 in \cite{2012A&A...537A.139C}, the expansion index
$m$ at the stagnation point changes quickly from a value of $\sim$0.8
to $\sim$0.45 during the first $\sim$30 yr after the encounter, while
the expansion in the opposite direction remains high.  Over the same
time frame the remnant radius in the direction of the stagnation point
grows more slowly so that the remnant starts to become asymmetric, but
the difference of the radii is not so large ($\lesssim$2\%). Thus this
model is consistent both with the large north-south variation of
expansion index observed in Kepler's SNR
\citep[$m=$~0.47--0.83:][]{2008ApJ...689..225K} and the modest
north-south radius difference as shown in Fig.~\ref{fig:cnt}
(right). 

In addition, we find that the bilateral protrusions at the
southeast/northwest rims match well with a simple elliptical geometry
(plotted as the yellow figures in Fig.~\ref{fig:cnt}).  In each panel
the ellipse is centered on the kinematic center for undecelerated
(left) or decelerated (right) knot motion and the axis lengths and
orientation are matched to the shape of the nonthermal filament on the
eastern rim.  Although the physical origin of these prominent
structures is not yet resolved, there have been suggestions that the
symmetric protrusions are related to the explosion
\citep[e.g.,][]{2013MNRAS.435..320T}.  If so, the notable agreement of
the shape and orientation of the protrusions to a simple elliptical
geometry centered on and symmetric about the kinematic center derived
from the decelerated knot motions, offers further support for this
being the site of the explosion.

\subsection{Spatial Density Variations in the Ambient Medium}

In one-dimensional SNR evolutionary models, a high value of the
expansion parameter indicates that the remnant is interacting with a
low density ambient medium. Here we derive an estimate of the density
required. \cite{1998ApJ...497..807D} investigated the dynamical
evolution of SN Ia assuming an exponential ejecta density profile.  An
expansion parameter $m \gtrsim$ 0.75 is realized in their models at a
scaled time of $t^{\prime} \lesssim$ 0.1 (see Fig.~2f in their paper).
The scaled time is related to the pre-shock ambient medium density
($n_{\rm H}$), and the remnant's age, explosion energy ($E_{51}$ in
units of $10^{51}$ ergs), and ejected mass ($M_{\rm ej}$) as
$$n_{\rm H} \approx 0.24\,  (t^\prime)^3\, E_{51}^{-3/2} (M_{\rm ej}/M_{\rm ch})^{5/2}\, {\rm cm}^{-3}$$
for the remnant's age of 401.7 yr.  Assuming typical values for the
explosion energy ($E_{51} = 1$) and ejected mass ($M_{\rm ej} = M_{\rm
  ch} = 1.4\, M_{\odot}$), allows us to convert the scaled time for
nearly undecelerated motion to an upper-limit on the ambient medium
density (assumed uniform) of $n_{\rm H} < 2.4 \times 10^{-4}$
cm$^{-3}$.  This value is comparable to the expected density at the
remnant's location above the Galactic plane in the absence of stellar
mass loss.

However, a compact knot, overdense with respect to its surroundings,
undergoes a considerably different type of evolution than does an
idealized spherically symmetric distribution of ejecta. To investigate
this scenario, we follow \citet{wangchevalier01} who simulated the
evolution of clumped ejecta in Tycho's SNR to understand the
conditions under which an ejecta knot could survive as it propagates
out to, and possibly deforms, the forward shock.  In this scenario a
compact ejecta knot enters the reverse shock at some time and
propagates outward through the high-pressure zone of shocked ejecta.
The impact of the reverse shock on the knot drives a transmitted shock
into the knot crushing it; in time this shock exits the knot which
sends a rarefaction wave through it.  Meanwhile instabilities develop
along the knot boundary due to shear flow and rapid local
accelerations that result in the destruction of the cloud after a few
cloud-crushing times.  This key timescale is given by $t_{\rm cc} =
\chi^{1/2} r_{\rm cloud}/ v_{\rm shock}$ \citep{1994ApJ...420..213K},
where $\chi$ is the density contrast of the knot with respect to the
intercloud medium, $r_{\rm cloud}$ is the initial radius of the cloud,
and $v_{\rm shock}$ is the velocity of the shock in the intercloud
medium.  Note that this timescale was originally developed for the
case of an interstellar cloud impacted by the forward shock of a SN
explosion, but is applicable to the closely analogous knot/reverse
shock situation we have here.

The survivability of an ejecta clump depends on its size (relative to
the size of the high pressure shocked ejecta zone) and density
contrast with respect to the rest of the ejecta.  Smaller
clumps with higher density contrast survive for longer.  Moreover the
drag on a high speed clump depends sensitively on the density
contrast, in the sense that a higher density contrast produces
relatively less drag.  \citet{wangchevalier01} therefore find that in
order for there to be undecelerated clumps of ejecta near the limb of
Tycho's SNR some 400 year after explosion, the density contrast needs
to be high, $\chi>100$.  

However, another option for increasing the survivability of an ejecta
knot is to have it interact with the reverse shock when that shock is
still forming during its early evolutionary phase. For example Fig.~8
of \citealt{wangchevalier01} shows a knot that survives and continues
expanding out to deform the forward shock from its initial interaction
with the reverse shock at a scaled time of $t^\prime = 0.217$ through
to $t^\prime \sim 0.8$.  The same scaling applies here as in the first
paragraph of this section so a value of the scaled time of $\sim$0.8
corresponds to an ambient medium density of $n_{\rm H} \sim 0.1$
cm$^{-3}$.  Such a low density would additionally allow for clumps
with lower density contrast to survive to the current age.  Thus, to
summarize, the presence of undecelerated knots in Kepler's SNR
requires either that those knots were generated with a high initial
density contrast ($\chi>100$) or that the ambient medium contains
lower-density ($n_{\rm H} \sim 0.1$ cm$^{-3}$) windows or gaps through
which potentially lower density contrast knots have propagated.

The evidence that Kepler's SNR is embedded in a dense environment is
strong.  Yet our work now suggests that the ambient medium could
be structured including both higher and lower density regions.
One possibility, as briefly explored by
\citet[][]{2013ApJ...764...63B}, might be that the donor star's wind
has sculpted a dense disk-like structure with lower densities
perpendicular to the disk plane.  Another option could rely on an
``accretion wind'' from the accreting white dwarf
\citep{1996ApJ...470L..97H}, which has the potential to blow a large,
low density cavity that allows for ejecta to expand rapidly
\citep{2007ApJ...662..472B}.  If the progenitor system to Kepler's SNR
had a bipolar outflow \citep[as in the case of the supersoft X-ray
  binary RX
  J0513.9--6951:][]{1993A&A...278L..39P,2002AJ....124.2833H}, the
ambient density along the polar axis could be much lower than
elsewhere.

\begin{figure}[h]
 \begin{center}
  \includegraphics[trim=18mm 51mm 2mm 40mm,clip,width=8cm]{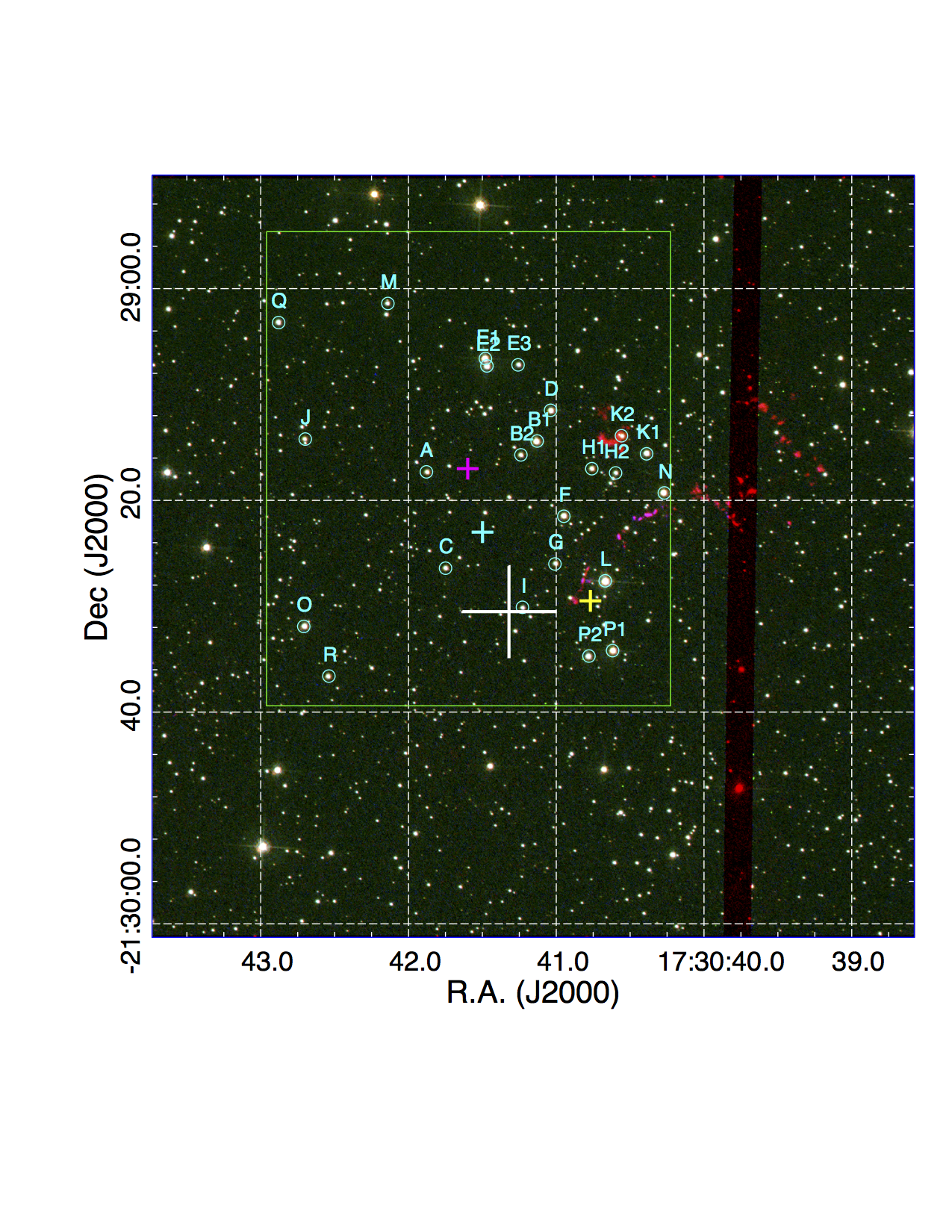}
 \end{center}
\caption{{\it Hubble Space Telescope} color image of Kepler's SNR
  using data from the Advanced Camera for Surveys (ACS) in the F502N
  (blue), F550M (green), and F660N (red) filters that trace [O III]
  $\lambda$ 5007 emission, the stellar continuum, and H$\alpha$,
  respectively. The green box is the region studied by
  \citet{2014ApJ...782...27K} where they measured spectra from the
  ground for a number of stars (not all of which were isolated).  The
  24 stars circled in cyan are those with V-band luminosities greater
  than $\sim$10$L_\odot$ at the distance of the remnant.  Published
  explosion centers are shown with small magenta, cyan, and yellow
  plus signs (same as in Fig.~\ref{fig:cnt}). The kinematic center reported here
  is shown as the large white cross, the size of which denotes the
  1$\sigma$ uncertainty.}
\label{fig:hst}
\end{figure}

\subsection{Implications for the Left-Over Companion Star}

Our new kinematic center allows us to reopen the search for a
surviving donor star under the SD scenario for the progenitor to
Kepler's SNR.  The most extensive study to date was carried out by
\citet{2014ApJ...782...27K}.  These authors identified two dozen stars
from {\it HST} with $V$-band luminosities greater than $\sim$10
$L_\odot$ assuming they lie at the distance of Kepler's SNR.  We
indicate these stars on Fig.~\ref{fig:hst} with cyan colored circles.
\citet{2014ApJ...782...27K} use ground-based optical spectroscopy to
measure radial velocities for these stars, although some of the
candidates (E, B, P, H, and K) were blended in the ground-based data, and
in addition it was not possible to obtain a reliable radial velocity
measurement for star I.

The radial velocities were compared to two velocity distributions: one
for field stars based on the Besan\c con model \citep{robin+03} of
galactic dynamics and the other based on the distribution of radial
velocities for SD donors ranging from main-sequence to giant stars
\citep{han08}. A probability was assigned to each star based on a
Monte Carlo simulation.  None of the stars were significant outliers
with respect to the Besan\c con model, although a number were
inconsistent with the expected donor distribution.
\citet{2014ApJ...782...27K} consider candidates E1, E2, K1, L, and N to
be the most notable for further follow-up, since they have $L > 20
L_\odot$ and a modest probability for being consistent with the
expected radial velocity distribution for a donor star.

We can now assess the probability of positional agreement between the
explosion center and each candidate.  The 3-$\sigma$ limit on the
allowed distance is 15$^{\prime\prime}$ to which we add
3.5$^{\prime\prime}$ to account for the donor's possible proper motion
($<$200 km s$^{-1}$).  Thirteen of the {\it HST} candidate stars fall
within this area, including stars L (the most luminous candidate with
$L_V = 86 L_\odot$), I (no radial velocity measurement), and G (second
highest donor probability based on radial velocity), although the
later two stars are of modest luminosity ($L_V = 7 L_\odot$ and
9$L_\odot$, respectively).  Although there remains no obvious donor
candidate for the traditional SD-scenario, our new kinematic center
has ruled out many of the interesting candidates suggested for
additional follow-up and has significantly reduced the search area for
donor candidates for modified SD-scenarios \citep[e.g.,][]{distefano+11,wheeler12}.

\section{Conclusions}\label{conclusion}

Most of our key results are based on the proper motion analysis of the
four available epochs (from 2000 to 2014) of \chandra\ ACIS-S X-ray
observations of Kepler's SNR.  We have discovered five X-ray--emitting
knots with no detectable optical emission  that are
moving with nearly undecelerated motion (i.e., with expansion indices
$>$ 0.75).  The proper motion of these knots extrapolate back, over
the age of the remnant, to a consistent and accurate center when their
(modest) deceleration is included. A number of prominent structures in
the remnant display a notable symmetry about the new kinematic center.
For example the similarity between the northern and southern radii
suggests that the forward shock of the remnant has encountered the
northern density enhancement fairly recently, within the last 100
years or so, as some models argue
\citep[e.g.,][]{2012A&A...537A.139C}.  The symmetric shape, extent and
orientation of the southeast/northwest protrusions about the kinematic
center add evidence to arguments that these protrusions may be related
to the explosion process itself.

Our spectral analysis provides information on the composition of the
knots. We report on three knots with near-solar abundances that show
little proper motion or radial velocity. The Ej knots were selected
because their associated H$\alpha$ shocks had measured proper motions
from {\it HST} data \citep{2016ApJ...817...36S}.  In the X-ray band
these knots have metal-enhanced abundances but with a larger abundance
of low-$Z$ species (O, Ne, Mg) compared to Si, S, and Fe than the
other X-ray ejecta knots studied here.  The Ej knots show high amounts
of deceleration and low 3D velocities. Beyond this, however, there is
no simple relationship between knot composition and motion.  The N
and NE series of knots share similar abundance patterns with high Si
and S abundances, some Fe, and very low O abundances but have a range
of expansion indices (from 0.45 to 0.95).  On the other hand the SW
knots show 3D speeds and expansion indices that fall between the NE
and N knots, but their spectra show considerably less enhancement of
the Si and S abundances compared to Fe.  This strongly indicates that
the origin of the knots (traced by composition) is independent of the
kinematics of the knots (traced by the expansion index). 

The measurement of radial velocities from spectral analysis of {\it
  Chandra} ACIS data is subject to systematic uncertainty from
detector (gain uncertainty) and analysis (background subtraction)
effects, so we summarize the key results from these measurements
separately here.  We find that the 3D space velocities of the highest
speed knots are in the range 9,100 km s$^{-1}$ $\lesssim v_{\rm 3D}
\lesssim$ 10,400 km s$^{-1}$, which is similar to the expansion speed
of Si-rich ejecta seen in the optical spectra of SN Ia near maximum
light.  We also find a correlation between 3D speed and expansion
index; the sense of the correlation is not surprising: higher speeds
correlate with higher expansion indices and vice versus.

We looked into the conditions that would allow for the existence of
high speed, undecelerated knots in Kepler's SNR some 400 years after
explosion, using the article by \citet{wangchevalier01} as a useful
guide.  One option would require that the knots formed with a high
density contrast (more than 100 times the density of the interknot
medium); another would require that knots with potentially lower
density contrast propagated through an ambient medium with a
relatively low density ($<$0.1 cm$^{-3}$). We favor the latter
interpretation since the generation of high-density-contrast clumps in
a SN Ia explosion seems less plausible to us than the possibility that
the environment of Kepler's SNR contains gaps or windows of lower
density gas.

Our new kinematic center has well-defined positional uncertainties
which have allowed us to refine the search for possible surviving donor
stars under the single-degenerate (SD) scenario for SN Ia.  As shown
before \citep{2014ApJ...782...27K} there are no viable candidates for
a traditional SD-scenario.  Nevertheless our new position for the
explosion center rules out several interesting candidates suggested by
others for further follow-up and has greatly reduced the area to be
searched for fainter donor stars under more exotic SD scenarios.

Our work has added new important pieces of evidence to the enigmatic
remnant of Kepler's SN that bear on both the nature of the explosion
and the structure of the ambient medium.  Further study of the
detailed composition of the new high-speed ejecta knots should allow
us to identify the conditions of the burning front where they formed
during the explosion. Mapping the positions and velocities of other
high-speed ejecta knots in Kepler's SNR should allow us to determine
the extent of the low density regions within the ambient medium. This
will be important information for understanding the progenitor system.

\acknowledgments{T.S.\ was supported by the Japan Society for the
  Promotion of Science (JSPS) KAKENHI Grant Number 16J03448.  This
  research was also supported in part by NASA grant NNX15AK71G to
  Rutgers University.  J.P.H.\ acknowledges the hospitality of the
  Flatiron Institute which is supported by the Simons Foundation.  We
  also thank Drs. Manabu Ishida, Yoshitomo Maeda, Ryo Iizuka, Saurabh
  Jha, Carles Badenes, and James Stone for helpful discussion and
  suggestions in preparing this article. }

\bibliographystyle{yahapj}

\end{document}